\pdfoutput=1 
\documentclass{amsart}
\usepackage{amsrefs}
\usepackage[utf8x]{inputenc}

\theoremstyle{definition}

\theoremstyle{remark}

\numberwithin{equation}{section}

\begin{document}

\title{Weyl Geometry and Quantum Corrections}


\author{Sijo K. Joseph}
\address{Department of Physics, GITAM (Deemed to be University), Hyderabad, India}
\email{skizhakk@gitam.edu}
\thanks{}



\date{\today}

\dedicatory{}

\begin{abstract}
Recent research in the geometric formulation of quantum theory has implied that Weyl
Geometry can be used to merge quantum theory and general relativity consistently as
classical field theories. In the Weyl Geometric framework, it seems that both quantum
theory and gravity can merge consistently, once quantum theory is geometrized. The
extended differential geometry can modify the quantum mechanical results into a more
general nonlinear framework. Author shows that, how the extended differential geometry
modifies the known quantum equations and also the modification to the Maxwell's
electromagnetic equations.
\end{abstract}

\maketitle

\section{Introduction}
It is well know that Einstein's theory of gravity is based on Riemann Geometry which is only a subclass of differential geometry. 
There are many mathematical extensions to Riemann geometry, for example the Weyl Geometry is one among them~\cite{WeylReview2017}. Einstein's gravity assumes a simple Lagrangian proportional to the curvature scalar ($\mathcal{L}=\sqrt{-g} R$ ). This is one of the simplest Lagrangians that one can assume. There are many possible generalizations of Einstein's theory, typical examples can be seen in these references ~\cite{BeyondEinsteinGravity,Bergmann1968,fR_HamFormlation,fofT1,fofT2,Bekenstein2011_TeVeS,
BekensteinPRD_TeVeS,Moffat_STVG,MOG_Moffat}.

Weyl had developed a purely infinitesimal geometry,  building upon a conformal generalization of the Riemannian metric $g_{\mu\nu}$, where a point-dependent rescaling is acheived via the the conformal factor $\Omega^2$
and it is given by $g_{\mu\nu}=\Omega^2 \bar{g}_{\mu\nu}$ where $\bar{g}_{\mu\nu}$ is our background metric.
It was F. Shojai et.al. ~\cite{Shojai2008} who discovered that the quantum theory can be incorporated into gravity in a geometric manner using the conformal factor $\Omega^2$. Conformal re-scaling  arises purely from a quantum mechanical quantity called the quantum potential~\cite{Shojai_Article,GabayJoseph1,GabayJoseph2,SKJoseph1}.  Such a geometrical unification of quantum theory to general theory of relativity is achieved using deBroglie-Bohm version of quantum theory~\cite{Bohm1975,BohmI,BohmII,Shojai_Weyl2003}.  Author is reviewing the Weyl Geometry and the correction terms appearing in scalar quantum mechanical wave-equation based on the previous studies~\cite{GabayJoseph1,GabayJoseph2,Joseph2018Geom}. We will simply focus on the mathematical corrections appearing in quantum mechanical equations as well as in the Maxwell's electromagnetic equations.

\section{Weyl Gravity Corrections in Klein-Gordon Equation}
In Riemannian geometry, the length of a vector remains constant and only the orientation changes during parallel transport . 
While in Weyl Geometry, both the length of a vector and its orientation changes during parallel transport. 
\begin {eqnarray}
\nabla_{\alpha}g_{\mu\nu} = 0 &\implies &\text{Riemannian Geometry} \\
\nabla_{\alpha}g_{\mu\nu} = \sigma_{\alpha} g_{\mu\nu} & \implies& \text{Weyl Geometry} 
\end{eqnarray}
In Riemannian geometry the covariant derivative of the metric is zero while it is not true in Weyl Geometry. Since the length changes during the parallel transport, the covariant derivative of the metric gives a nonzero contribution.

Weyl geometry encountered a serious criticism by Einstien, he criticized  that the clock will tick based on its history and different clock with different history will click at different rates which is not observed in reality hence Weyl theory cannot be correct. That is a criticism posed by Einstein in order to reject Weyl Geometry. This criticism can be easily overcome by adopting the Integrable Weyl Geometry. Weyl integrable geometry is given by the following condition,
\begin{equation}
\sigma=d\phi,
\end{equation}
here $\phi$ is a scalar field. Once $\sigma$ is a perfect differential of a scalar field, the integral over a closed curve vanishes. Using this argument, the criticism by Einstein can be circumvented, if the clock is completed a closed route, then the integral vanishes, and the integral is no-longer path dependent. Hence in integrable Weyl geomety, Einstein's criticism is no longer valid. In Weyl geometry, one can think about different conformal frames where the metric is different by a conformal scaling but the Physics remains the sames.

Here the following equation i.e. Eq.~\ref{WeylFrames} gives a different description of the same manifold with a conformally re-scaled metric and Weyl field $\sigma$.
\begin{equation}
(M,g,\sigma)\to (M,\bar{g},\bar{\sigma}) \label{WeylFrames}.
\end{equation}

The different conformal frames are related by the following transformation equations and it is called Weyl Gauge transformation,
\begin {eqnarray}
\bar{g}={e}^{-f} g\\
\bar{\sigma}=\sigma-df
\end{eqnarray} where $f$ is a scalar function on the manifold $M$. Let us concentrate on the conformal transformation of the metric in different Weyl frames.
\begin{equation}
(M,g,\sigma)\to (M,\bar{g},\bar{\sigma})
\end{equation}
\begin {eqnarray}
\bar{g}={e}^{-f} g
\end{eqnarray}
Different Weyl frames are related via conformal transformations yet it describe the same Physics. This concept can be easily understood, once we focus on the geometric formulation of quantum theory using conformal transformation.

\section{Geometric Formulation of Quantum Theory}
Here our aim is to show that quantum theory is associated to the conformal frames as described in Weyl Geometry. 
Consider the Lagrangian density of Klein-Gordon Field, which  is given by,
\begin{eqnarray}
\mathcal{L}=\frac{1}{2}\partial_{\mu}\Phi^{*}\partial^{\mu}\Phi-\frac{1}{2}\frac{m^2}{\hbar^2}\Phi^{*}\Phi
\end{eqnarray}
Euler-Lagrange equation gives,
\begin{eqnarray}
&\mathcal{S} = \int{\mathcal{L}\sqrt{-g}\, \mathrm{d}^4x}. \\
&\frac{\delta \mathcal{S}}{\delta\phi}=0 \implies \frac{\partial\mathcal{L}}{\partial\phi} -\partial_\mu  \left(\frac{\partial\mathcal{L}}{\partial(\partial_\mu\phi)}\right)=0 
\end{eqnarray}
Hence the  Klein-Gordon Equation is obtained using standard quantum theory. 
\begin{eqnarray}
(\partial_{\mu}\partial^{\mu}+\frac{m^2}{\hbar^2})\Phi=0 \label{KGEq}
\end{eqnarray}
Note that, once can use a complex field to describe the quantum theory. Gravity has a highly different mathematical framework which uses real quantities and differential geometry is utilized to describe the physical mechanism. In quantum theory, we  adopt complex fields on a flat Minkowski space-time. In order to geometrize quantum theory, we need to return to the real field variables, hence decomposing the complex field into two real field equation, one can geometrize quantum theory. Taking $\Phi=\sqrt{\rho}{e}^{\frac{i}{\hbar}S}$, we will be able to decompose the Klein-Gordon equation into two real field equations. Substituting polar form of $\Phi$  in Eq.~\ref{KGEq} and separating real and imaginary part we get two different field equations.

\begin{eqnarray}
\partial_{\mu}S\partial^{\mu}S&=&m^2\Omega^2 \label{EqMo1}\\
\partial_{\mu}(\rho\partial^{\mu}S)&=&0 
\end{eqnarray}
where $\Omega^2=1+\frac{\hbar^2}{m^2}\frac{\partial_{\mu}\partial^{\mu}\sqrt{\rho}}{\sqrt{\rho}}$. One can notice that $\Omega^2$ looks like a conformal factor here. If we start with classical equations and make a conformal transformation, one can get the quantum mechanical equations.  This is a differential geometric way of finding quantum equations.  Once can see that with a proper conformal scaling of the Minkowski space, one can reach into classical equation of motion. For example, Eq.~\ref{EqMo1}, can be re-casted into differential geometric form by dividing it with $\Omega^2$.
\begin{eqnarray}
\frac{1}{\Omega^2}\eta^{\mu\nu}\partial_{\mu}\partial_{\nu}S&=&m^2 \\
g^{\mu\nu}\partial_{\mu}S\partial_{\nu}S&=&m^2 \label{EqMotion1}
\end{eqnarray}
Here the new metric $g_{\mu\nu}$ can be written as a conformal transformation to the Minkowski metric  ${\eta_{\mu\nu}}$.
\begin{eqnarray}
g^{\mu\nu}=\frac{\eta^{\mu\nu}}{\Omega^2} \implies g_{\mu\nu}={\Omega^2} {\eta_{\mu\nu}}
\end{eqnarray}
Here  Eq.~\ref{EqMotion1} obtained looks like a free-particle classical equation motion on a curved manifold with metric $g_{\mu\nu}={\Omega^2} {\eta_{\mu\nu}}$. Hence a conformal re-scaling of Minkowski space-time $\eta_{\mu\nu}$ makes the quantum equations into classical equations of motion. In other words, quantum theory can be absorbed into a conformally transformed metric. Treating the space-time  as conformally curved and exploring the classical motion on this manifold is theoretically equivalent to studying a quantum problem on a flat Minkowski space-time. This opens the freedom to choose different conformal frames. For example when the background metric is curved due to gravity ie $\bar{g}_{\mu\nu}$, we can consider a re-scaled metric ${g}_{\mu\nu}=\Omega^2\bar{g}_{\mu\nu}$. There are many ways to choose this total metric ${g}_{\mu\nu}$, hence there are different conformal frames which can give same Physics as long as the total metric is the same. These conformal frames corresponds to different ways of mixing quantum theory and Einstein gravity. Hence incorporating quantum theory in differential geometric framework one is naturally compelled to extend differential geometry in a much wider mathematical framework. One can associate gravity to the the coordinate transformation while quantum theory is associated with the length change during parallel transport of a vector. Hence Weyl Geometry is a mathematical framework to hold both quantum theory and Einstein's gravity.

\section{Coupling Gravity with Geometrized Quantum Theory}
Let us begin with the following classical action, 
\begin{eqnarray}
A[g_{\mu\nu},S, \rho, \lambda]&=& {\frac{1}{2\kappa}\int{d^4x\sqrt{-g} R}}\nonumber\\
 & &{+\int{d^4x\sqrt{-g} \left(\frac{\rho}{m}\nabla_{\mu}S \nabla^{\mu}S-m\rho\right)}}
 \label{Actioneq1} 
\end{eqnarray}
In order to incorporate the quantum effect, one need to make a conformal transformation to the aforementioned action and one can get,
\begin{eqnarray}
& & A[g_{\mu\nu},{\Omega^2}, S, \rho, \lambda]= {\frac{1}{2\kappa}\int{d^4x\sqrt{-g}\left(R\Omega^2-6\nabla_{\mu}\Omega\nabla^{\mu}\Omega\right)}}  \nonumber \\
& & {+\int{d^4x\sqrt{-g} \left(\frac{\rho}{m}\Omega^2 \nabla_{\mu}S \nabla^{\mu}S-m\rho\Omega^4\right)}}
 \nonumber \\
& &{+\int{d^4x\sqrt{-g}\lambda\left[\ln{\Omega^2}-\left(\frac{\hbar^2}{m^2}\frac{\nabla_{\mu}\nabla^{\mu}\sqrt{\rho}}{\sqrt{\rho}}\right)\right]}} \label{Actioneq} .
\end{eqnarray}
In order to get correct quantum mechanical equations, one needs to impose a constraint condition which is done in the third term   using a Lagrange multiplier $\lambda$.  The constraint equation is given by, 
\begin{eqnarray}
{\Omega^2 = \exp\Bigl(\frac{\hbar^2}{m^2}\frac{\nabla_{\mu}\nabla^{\mu}\sqrt{\rho}}{\sqrt{\rho}}}\Bigr)
\approx 1+\frac{\hbar^2}{m^2}\frac{\nabla_{\mu}\nabla^{\mu}\sqrt{\rho}}{\sqrt{\rho}} \label{ConstraintEq}
\end{eqnarray}
Minimizing the action, one is lead to generalized equations for the real-fields $S$ and $\rho$ and it is given by,

\begin{eqnarray}
{(\nabla_{\mu}S \nabla^{\mu}S- m^2\Omega^2)+\frac{\hbar^2}{2m\Omega^2 \sqrt{\rho}}[\Box({\frac{\lambda}{\sqrt{\rho}})}-\lambda\frac{\Box\sqrt{\rho}}{\rho}]=0}  \label{EqMotion}
\end{eqnarray}
\begin{eqnarray}
{\nabla_{\nu}(\rho\Omega^2\nabla^{\nu}S) =0} \label{ContinuityEq} 
\end{eqnarray}
Here Eq.~\ref{EqMotion} and Eq.~\ref{ContinuityEq} together gives a generalized Klein-Gordon Equation once seen in the wavefunction picture $\Phi=\sqrt{\rho}{e}^{\frac{i}{\hbar}S}$. Even in the flat space-time $\lambda$ correction exists in Eq.~\ref{EqMotion}. Combining Eq.~\ref{EqMotion} and Eq.~\ref{ContinuityEq}, one can get the complex quantum equation which is a generalized  Klein-Gordon equation with source and dissipation terms.
\begin{eqnarray}
{\Box \Phi +\frac{i}{\hbar}\frac{\nabla_{\mu}\Omega^2}{\Omega^2}\nabla^{\mu}S\Phi+\frac{m^2}{\hbar^2}\Phi=\frac{1}{2m\Omega^2 \sqrt{\rho}}[\Box({\frac{\lambda}{\sqrt{\rho}})}-\lambda\frac{\Box\sqrt{\rho}}{\rho}]\Phi} 
\end{eqnarray}
It simultaneously satisfy the following $\lambda$ equation (Eq.~\ref{lambda_deq2}),
\begin{eqnarray}
{\lambda= \frac{1}{(1-Q)}\frac{\hbar^2}{m^2}\nabla_{\sigma}(\lambda\frac{\nabla^{\sigma}{\sqrt{\rho}}}{\sqrt{\rho}})} \label{lambda_deq2}.
\end{eqnarray}
Hence we are lead to a more general quantum theory with the presence of a dissipation and source contribution. 
Taking $\lambda=0$ or $ \lambda=\rho$, one can get the complex Klein-Gordon equation without the source term. 
\begin{eqnarray}
{\Box \Phi +\Bigl(\frac{i}{\hbar}\frac{\nabla_{\mu}\Omega^2}{\Omega^2}\nabla^{\mu}S\Bigr) \Phi+\frac{m^2}{\hbar^2}\Phi=0}
\end{eqnarray}
If we assume that quantum force coupling contribution ($\Bigl(\frac{i}{\hbar}\frac{\nabla_{\mu}\Omega^2}{\Omega^2}\nabla^{\mu}S\Bigr) \Phi$ ) is negligible then one can recover standard Klein-Gordon Equations.
\begin{eqnarray}
{\Box \Phi +\frac{m^2}{\hbar^2}\Phi}=0
\end{eqnarray}

\section{Gravitational Corrections in Maxwell's Equations}
In order to couple electro-magnetism with matter field equation, we will simply add the Maxwell's Lagragian and the matter current is modified accordingly. The following action is proposed for the electromagnetic coupling (see~\cite{Joseph2018Geom} for more details),
\begin{eqnarray}
& &A[g_{\mu\nu},{\Omega}, S, \rho, A_\mu,\lambda]=\frac{1}{2k}\int{d^4x\sqrt{-g}\left(R\Omega^2-6\nabla_{\mu}\Omega\nabla^{\mu}\Omega\right)}\nonumber\\
& &+\int{d^4x\sqrt{-g} \left(\frac{\rho}{m}\Omega^2 (\nabla_{\mu}{S}+e A_{\mu})(\nabla^{\mu}{S}+e A^{\mu})-m\rho\Omega^4\right)} \nonumber\\
& &-\frac{1}{4}\int{d^4x\sqrt{-g}\, F_{\mu\nu} F^{\mu\nu}}  \nonumber\\
& &+\int{d^4x\sqrt{-g}\lambda 
\left[\ln(\Omega^2)-\left(\frac{\hbar^2}{m^2}\frac{\nabla_{\mu}\nabla^{\mu}\sqrt{\rho}}{\sqrt{\rho}}\right)\right]},
\end{eqnarray}

Maxwell's equation can be found by varying  the action with respect to $A_{\mu}$,
\begin{eqnarray}
\nabla_{\mu}F^{\mu\nu}-\frac{2e}{m}\rho\Omega^2(\nabla^{\nu}S+e A^{\nu}) =0. \label{GenMaxwellEq}  
\end{eqnarray}

Continuity equation is obtained from the variation of action with respect to $S$. This is just the imaginary part of the
quantum gravity corrected Klein-Gordon-Maxwell equation.
\begin{eqnarray}
\nabla_{\mu}\Bigl(\rho\Omega^2(\nabla^{\mu}S+e A^{\mu})\Bigr)=0 \label{GenEMContiEq}
\end{eqnarray}

In Eq.~\ref{GenMaxwellEq} we get a more generalized Electromagnetic equation. Let us consider the flat space-time and see how this equation behaves in a Minkowski space-time, then one can get
\begin{eqnarray}
\partial_{\mu}F^{\mu\nu}-\frac{2e}{m}\rho\Omega^2(\partial^{\nu}S+e A^{\nu}) =0. \label{GenMaxwellEq}  
\end{eqnarray}
Rewriting the expression of $F^{\mu\nu}$ in terms of the electro-magnetic vector potential $A^{\nu}$ in lorentz gauge, one can see that
\begin{eqnarray}
\Bigl(\Box -\frac{2e^2\rho}{m}\Omega^2\Bigr) A^{\nu}   =\frac{2e}{m}\rho\Omega^2\partial^{\nu}S. \label{GenMaxwellFlatEq}  
\end{eqnarray}
Note that quantum mechanical current is simply $J_{qm}^{\nu}=\rho\partial^{\nu}S$ while we have a conformally corrected quantum current density $J'^{\nu}=\Omega^2\rho\partial^{\nu}S$.Once $\Omega^2=e^Q$ is taken into account and higher order terms are neglected the total current density can be expressed as $J'^{\nu}\approx J_{qm}^{\nu}+Q J_{qm}^{\nu}$. Then the generalized Maxwells equations can be written as

\begin{eqnarray}
\Bigl(\Box -\frac{2e^2\rho}{m}(1+Q)\Bigr) A^{\nu}   =\frac{2e}{m}\Bigl( J_{qm}^{\nu}+Q J_{qm}^{\nu}\Bigr), \label{GenMaxwellFlatEq2}  
\end{eqnarray}
where $Q$ is the quantum potential. One can see that Maxwell's equation is corrected with a Proca like mass term with negative signature. Similarly the quantum mechanical current contains a contribution from the Conformal factor. Note that the conformal factor is given by $\Omega^2=e^Q$, Eq.~\ref{GenMaxwellFlatEq2} is found by ignoring higher order $\hbar$ terms, and one can see that in Eq.~\ref{GenMaxwellEq}, when the quantum potential $Q\to0$ one can recover the massless Maxwell's equation.

\section{Conclusions}
In this article we have seen that quantum mechanical equation contains a dissipation and source term correction due to the coupling between gravity and quantum theory. Similarly in electromagnetism a mass term and an extra conformal  current term appears as the gravitational correction. The coupling is not only due to the curved background geometry, even in flat space, the correction terms appears even though it is a small quantity. For all the practical purposes, we can deal with the linear partial differential equation while dealing with the quantum theory but this is only a special case of the gravitationally corrected rich nonlinear quantum equations.


\end{document}